\documentclass[]{aastex701}
\usepackage{soul}
\usepackage{color}

\newcommand{\SSFR}{$\Sigma_{\rm SFR}$}
\newcommand{\SP}{$\Sigma_{\rm SFR}-P$}
\newcommand{\Sg}{$\Sigma_{\rm g}$}

\begin{document}

\title{Pressure and Star Formation in LITTLE THINGS Dwarf Irregular Galaxies}

\author[0000-0002-1723-6330]{Bruce G. Elmegreen}\affiliation{Katonah, NY 10536}
\email[show]{belmegreen@gmail.com}  

\author[0000-0002-3322-9798]{Deidre A. Hunter}\affiliation{Lowell Observatory, 1400 West Mars Hill Road, Flagstaff, AZ 86001, USA}
\email{dah@lowell.edu}  

\author[0000-0002-5788-2628]{Edvige Corbelli} \affiliation{INAF-Osservatorio Astrofisico di Arcetri, Largo E. Fermi, 5, 50125 Firenze, Italy}
\email{edvige.corbelli@inaf.it} 
\begin{abstract}
The surface densities of star formation, \SSFR, in 24 dwarf irregular (dIrr) galaxies from the LITTLE THINGS survey are combined with gas surface densities and midplane pressures to examine the correlations found previously for spiral galaxies. The pressure is the weight of the disk inside the gas layer, including gas, stars, and dark matter, which usually dominates disk gravity in dIrrs. We compare the results to the outer part of M33, which has similar local properties but a slightly higher metallicity, enabling the detection of CO. All the data are convolved to the HI beam, but to study the effects of resolution, the galaxies are examined first with average radial profiles, and then with maps having $1.5^{\prime\prime}$ pixels and 244 pc pixels. The correlations are found to be independent of resolution from 24 pc to 424 pc.  The average ratio of molecular to atomic surface density is estimated to be $0.23\pm0.1$, from the H$_2$ surface density in M33 compared to the HI surface density at the same \SSFR\ in the dIrrs. With this ratio, the average star formation rate per molecule is about the same for all the dIrrs, and a factor of 2 less than the rate in M33. The pressure in dIrrs is so low that CO is essentially a dense gas tracer, with the same surface density threshold at the low metallicities of dIrrs as HCN has in spiral galaxies. As a result, CO regions in dIrrs should be strongly self-gravitating. 
\end{abstract}

\section{Introduction}
The surface density of star formation, \SSFR, correlates well with the total midplane pressure, $P$, in galaxy surveys such as THINGS \citep{leroy08}, KINGFISH \citep{herrera17}, DYNAMO \citep{fisher19}, PHANGS \citep{sun20}, EDGE-CALAFA \citep{barrera21}, ALMaQUEST \citep{ellison24}, and PHANGS-MeerKAT with MHONGOOSE \citep{eibensteiner24}. \cite{gallagher18} show a similar correlation between the star formation rate per unit dense gas and $P$. 
In 34 PHANGS galaxies, kpc-scale variations in \SSFR\ and $P$ relative to their radial averages also correlate well, and with the same slope as the correlation among total quantities, suggesting \SSFR\ follows $P$ on the $\sim50$ Myr timescale of star formation \citep{elmegreen24}. 
The correlation also holds for all $30^{\prime\prime}$ (120-pc) regions in the optical part of the galaxy M33, where it is even better after
removing a secondary correlation with the gas velocity dispersion, which is systematically lower at higher \SSFR\ for a given $P$ \citep{corbelli25}.

The correlations shown by these studies depend on the definition of pressure and on the values assumed for the unobserved quantities, such as the stellar scale height, which enters into the midplane stellar density, and the magnetic and cosmic ray pressures. Dark matter also contributes to the pressure in the disk. Radial variations of these unobserved quantities affect the slope of the correlation. \cite{ellison24} and \cite{corbelli25}, with slightly different expressions for $P$, use random forest regressions to show that the \SP\ correlation is tighter than the correlation between \SSFR\ and gas surface density, $\Sigma_{\rm g}$, which is the ``Kennicutt-Schmidt'' relation, hereafter KS \citep{kennicutt12}, for either molecules or total gas. 

The EDGE-CALAFA survey contains 96 galaxies, including some low-metallicity dwarf irregulars (hereafter dIrrs) that have a factor of $\sim3$ shift toward lower P for their \SSFR\  \citep{barrera21}. 
Here we look at local dIrrs to consider whether this shift might be the result of dark gas, which causes the pressure to be underestimated.  We examine the \SP\ relation for 24 galaxies from the Local Irregulars That Trace Luminosity Extremes The HI Nearby Galaxy Survey (LITTLE THINGS), where we have sufficient data to determine midplane pressures \citep{hunter12}. This is the same sample as that studied previously for the radial dependence of the relationship between \SSFR\ and HI gas surface density \citep{elmegreen15}, which is considered here also. 

The dIrr galaxy WLM, at a distance of 985 kpc and a metallicity of $0.13\;Z_\odot$ \citep{lee05,asplund09,leaman12} offers a good example of dark gas.  There is a collection of small ($\sim 2$ pc) CO clouds in a northern region (``Region A''), surrounded by dust emission observed at $160\mu$m and $870\mu$m and HI emission \citep{elmegreen13,rubio15,cigan16}.  There are other collections of small CO clouds elsewhere too \citep{archer24,archer25}. The stellar surface density can be estimated from the surface brightness and colors, and the dark matter density from abundance matching with the observed stellar mass (see below).  Using the HI velocity dispersion and a reasonable estimate for the stellar dispersion, the pressure in the northern region is $\sim1.2\times10^5 k_{\rm B}$ (see below). For the \SP\ relationships in the above references, \SSFR\ at this pressure would be in the range from $\sim3.1\times10^{-3}$ \citep{barrera21} to $\sim5.6\times10^{-2}$ \citep{herrera17}, with an average of $\sim2\times10^{-2}$
(in units of $M_\odot$ pc$^{-2}$ Myr$^{-1}$). This is close to the observed \SSFR\ in this region, which is $6.7\times10^{-3}$ from H$\alpha$ and $8.3\times10^{-3}$ from FUV \citep{elmegreen13}.
The total gas surface density, $58\;M_\odot$ pc$^{-2}$, has been derived from the dust emission, and the atomic gas surface density, $27.3\;M_\odot$ pc$^{-2}$, has been derived from HI emission. Together, these imply a ratio of dark gas (the difference between these two) to atomic gas of 1.1 in this small region.

The \SP\ relationship for dIrrs is interesting because these galaxies appear to have weakly gravitating disks. In annular averages, the conventional Toomre Q values for the gas and star components (assuming thin disks) are in the range $\sim1-10$ and $\sim1-50$, respectively, and they are approxmately twice larger for more realistic thicknesses \citep{elmegreen15}. Although it is possible that dark gas makes $Q$ closer to unity \citep{hunter19}, the lack of stellar spirals also suggests the disks are fairly stable \citep[see also][]{hunter98}. Also in dIrrs, the KS relation is steeper than in spiral galaxies, meaning the SFR is low for its amount of gas \citep{vanzee01,leroy08,roy09,roy14}. If the \SP\ slope for dIrrs is about the same as for spirals \citep[e.g.,][]{barrera21}, but the KS slope for dIrrs is steeper, then the \SP\ law could be more fundamental. 

dIrr galaxies are also interesting because they  have relatively more dark matter than stars and gas compared to spiral galaxies \citep{carignan88,deblok97,oh15}, and they have more gas relative to stars \citep{leroy08,bradford15,elmegreen15}. Gas, stars, and dark matter all influence the \SP\ relation, but only gas is involved with the KS relation.  These differences in the relative contributions to $P$ for the two galaxy types could reveal if all three components in $P$ are necessary to predict \SSFR, or if gas alone dominates. 

The stellar contribution to the \SP\ correlation resembles the {\it extended} KS relation, where stellar surface density is included \citep{shi11}. In a common formulation for the dynamical pressure \citep{ostriker22}, the first term contains a power law of the gas surface density and is similar to the KS relation, and the second term contains both gas and stars and is similar to the extended KS relation. These similarities suggest that the combined terms, which represent pressure, are the composite of the two KS relationships and not fundamentally different \citep{elmegreen24}.

In what follows, we determine the interstellar pressure and the gas and star formation surface densities for the 24 dIrrs in three ways: (1) using azimuthally-averaged quantities; (2) using values determined in each pixel on a $1.5^{\prime\prime}$-square grid, and (3) with quantities determined in each pixel on a 244 pc-square grid. Because these dIrrs are irregular in their distributions of gas and star formation, the azimuthally averaged quantities are often different from the local quantities determined inside pixels at the same radius. The two pixellated versions illustrate a key result that the KS and SP relations are essentially independent of physical resolution on scales below a few hundred pc. Smaller pixels have more scatter in the relationships but approximately the same mean trends. 

The value of 244 pc is chosen because that is the pixel scale for most of the analysis of M33 in \cite{corbelli25} and because this is close to the lowest physical resolution for HI in our sample. To gain some insight on the molecular content of the dIrrs, where CO is usually not observed, we compare the KS and \SP\ relations for dIrrs to those in the outer part of M33. M33 is a low-mass, local group \citep[840 kpc distance;][]{gieren13} spiral galaxy, with an outer part that resembles a dIrr in several respects. Beyond $\sim4$ kpc radius, which is $\sim2$ exponential scale lengths \citep{verley09}, the surface density of HI is $\sim10\;M_\odot$ pc$^{-2}$ in M33, which is comparable to that in the central regions of dirrs, and it is larger than the surface density of stars, which is also the case throughout dIrrs \citep{elmegreen15,corbelli25}. Also in M33, the midplane gas density in the outer part is larger than the midplane stellar density, as in most parts of dIrrs. The HI layer in the outer regions of M33 is relatively thick too, with a scale height larger than 300 pc \citep{corbelli25}, as in dIrrs \citep{elmegreen15}. Also in both, there are no strong stellar density waves. The outer part of M33 between 4 and 8 kpc radius has an oxygen abundance, $12+\log(O/H)$, between 8.322 and 8.146, which is between 46\% and 31\% solar \citep{magrini10}.  This is consistent with a lower dust-to-gas ratio in the outer part of M33 \citep{corbelli25}. It is a factor of $\sim2$ or more higher than the oxygen abundance in dIrrs, which means that CO can be detected slightly more easily in M33. In dIrrs, the molecular or cold atomic gas associated with star formation is usually dark \citep{hunter24,park25}. The molecule-to-atom ratio in the outer part of M33 is $\sim15$\% \citep{corbelli25}, although it could be larger if there is unknown dark gas. 

\section{Galaxy Data}

The 24 LITTLE THINGS galaxies considered here are listed in Table 1 \citep[see also][]{elmegreen15}. The star formation rates are from GALEX far-UV (FUV) emission \citep{martin05}\footnote{GALEX was operated for NASA by the California Institute of Technology under NASA contract NAS5-98034}, the stellar mass surface densities are from SED fits \citep{zhang12} using GALEX FUV and near-UV, U, B, V and H$\alpha$ filters from the LITTLE THINGS survey \citep{hunter12},  and $3.6\mu$m images from the Spitzer Space Telescope \citep{fazio04}. A  Chabrier IMF is assumed. The HI surface densities, corrected to face-on orientation and multiplied by 1.36 to account for He and heavy elements, and the HI gas velocity dispersions, are from our VLA\footnote{The VLA is a facility of the National Radio Astronomy Observatory (NRAO), which is a facility of the National Science Foundation operated under cooperative agreement by Associated Universities, Inc. These data were taken during the upgrade of the VLA to the Expanded VLA, now JVLA.} observations \citep{hunter12}. The HI sensitivity is better than 1.1 mJy beam$^{-1}$ per channel. All of the maps have been smoothed to the spatial resolution of the HI, which varies between $5^{\prime\prime}$ and $16^{\prime\prime}$ or between 24 pc and 424 pc, depending on distance. The stellar velocity dispersions are assumed to be constant with radius and are evaluated from the expression $c_s=10^{-0.15M_B-1.27}$ km s$^{-1}$ \citep{swaters99,johnson12,johnson15} for B-band absolute magnitude $M_B$. 

Dark matter is an important component in the disks of dIrr galaxies \citep{oh11}. The dark matter density can be estimated from the dynamical analysis of rotation curves or by using abundance matching, which relates simulation results to the galaxy luminosity distribution function. For the latter method, \cite{corbelli25} give an explicit expression for the dark matter density as a function of radius for spiral galaxies using NFW \citep{navarro97} profiles. For dwarf galaxies examined in this paper, halo masses are expected to be in the range of $10^{10}\;M_\odot$ to $10^{11}\;M_\odot$ where cored dark matter haloss are more likely to form \citep{dicintio14}. Therefore, for abundance matching density estimates we assume a cored density profile as given by \cite{burkert95},
\begin{equation}
\rho_{\rm DM}(r) = {{\rho_0 r_0^3}\over{(r+r_0)(r^2+r_0^2)}},
\label{DM1}
\end{equation}
where $\rho_0$ and r$_0$ are the core density and radius. We write the virial mass, the mass of a galaxy's halo within a radius where the average density is 200 times the critical density of the universe, as a function of the core density using a Hubble constant of 70 km s$^{-1}$ Mpc$^{-1}$, and then the core radius satisfies
\begin{equation}
r_0 = 63/\rho_0
\label{DM2}
\end{equation}
with central density in units of $M_\odot$ pc$^{-3}$ and r$_0$ in pc. Equation (\ref{DM2}) implies the central surface density of dark matter is constant \citep{kormendy04}, $\log\rho_0r_0=1.8$ for units of $M_\odot$ pc$^{-2}$. This is essentially the value found by \cite{kormendy16} for a wide range galaxy luminosities, slightly smaller than in \cite{donato09}, who find $\log\rho_0r_0=2.15\pm0.2$ for all luminosities, but consistent with the pseudo-isothermal halo models tested by \cite{li19} for dwarfs, who show a slightly decreasing trend at lower luminosities for halos with cores. This dark matter core surface density is also predicted by some dark matter models for halo masses of order 10$^{11}$~M$_{\sun}$ \citep{lin16}. The halo mass was connected to the stellar mass using an abundance matching technique from cosmological simulations in \cite{dicintio14}, who also consider how feedback affects the halo cusp-to-core transition.  The final relation between the core dark matter density in units of $M_\odot$ pc$^{-3}$ and the stellar mass in units of $M_\odot$ which we use in this paper reads:
\begin{equation}
\log\rho_0=-0.175-0.21\log M_{\rm stars} 
\label{DM3}
\end{equation}

The stellar masses are given in Table 1. They are based on photometry in \cite{hunter06}, using $M_{\rm V}$ with mass-to-light ratio from $B-V$ as in \cite{herrmann16}. The extinction within the dIrrs assumed $E(B-V)=0.05$, which was added to the foreground value that is usually a few hundreths, except for NGC 1569. The B-V value used the largest ellipse for the photometry, including essentially the whole optical galaxy. 

An alternate derivation for the halo mass profile is in \cite{oh15}, based on the observed HI rotation curve and stellar mass profile, and assuming an inclination from the observed ratio of axes.  All of the pressures discussed here were evaluated with both the Oh et al dark matter profiles and equations (\ref{DM1}) to (\ref{DM3}). There were some discrepancies, particularly for IC 1613 and NGC 3738, where the \cite{oh15} halo masses are factors of $\sim3$ lower and higher, respectively, than the others scaled to their stellar masses.  These discrepancies could result from incorrect inclinations \citep{oman16}.

\begin{center}
\begin{deluxetable}{lccclccc}
\tablenum{1} \tablecolumns{8} \tablewidth{315pt} \tablecaption{The Galaxy Sample
\label{tab-sample} } 
\tablehead{
\colhead{Galaxy} & 
\colhead{D} & 
\colhead{$M_{\rm B}$} & 
\colhead{$M_{\rm stars}$} & 
\colhead{Galaxy} &
\colhead{D} & 
\colhead{$M_{\rm B}$} & 
\colhead{$M_{\rm stars}$} \\
\colhead{} & 
\colhead{(Mpc)} & 
\colhead{(mag)} & 
\colhead{($\times10^7\;M_\odot$)} &
\colhead{}&
\colhead{(Mpc)} & 
\colhead{(mag)} & 
\colhead{($\times10^7\;M_\odot$)}
}
\startdata
CVnIdwA   &  3.6 & -12.2 & $0.283\pm0.059$ & DDO 154 & 3.7 & -13.9 & $1.91\pm0.06$ \\
DDO 43    & 7.8  & -14.6 & $4.35\pm0.16$  & DDO 168 & 4.3 & -14.7 & $9.20\pm0.08$ \\
DDO 46    & 6.1  & -14.1 & $3.01\pm0.09$  & DDO 210 & 0.9 & -10.4 & $0.147\pm0.002$\\
DDO 47    & 5.2  & -15.2 & $5.98\pm0.15$  & DDO 216\tablenotemark{a} & 1.1 & -13.1 & $2.97\pm0.01$ \\
DDO 50    & 3.4  & -16.4 & $14.3\pm0.1$   & F564-V3 & 8.7 & -13.6 & $1.84\pm0.09$ \\
DDO 52    & 10.3 & -15.1 & $7.51\pm0.41$  & Haro 29 & 5.8 & -14.0 & $2.63\pm0.03$ \\
DDO 53    &  3.6  & -12.6 & $1.78\pm0.02$ & Haro 36 & 9.2 & -15.5 & $11.4\pm0.01$ \\
DDO 70    &  1.3  & -13.7 & $2.01\pm0.01$ & IC 1613 & 0.7 & -14.2 & $3.76\pm0.02$ \\
DDO 87    &  7.7  & -14.5 & $5.69\pm0.19$ & NGC 1569 & 3.4 & -17.9 & $69.0\pm0.02$ \\
DDO 101   &  6.4 & -14.4 & $8.54\pm0.08$  & NGC 2366 & 3.4 & -16.1 & $20.6\pm0.04$ \\
DDO 126   &  4.9 & -14.5 & $3.37\pm0.08$  & NGC 3738 & 4.9 & -16.3 & $37.5\pm0.1$ \\
DDO 133   &  3.5 & -13.8 & $3.90\pm0.06$  & WLM      & 1.0 & -14.0 & $2.96\pm0.02$ \\
\enddata
\tablenotetext{a}{The pixel size for all original HI maps is $1.5^{\prime\prime}$ except for DDO 216, where it is $3.5^{\prime\prime}$.}
\end{deluxetable}
\end{center}

\section{Pressure}
In the gas disk of a galaxy, the average equilibrium midplane pressure balances the perpendicular gravitational force per unit area on the gas. There are many formulae for this pressure in the literature \citep{elmegreen89,boulares90,wong02,blitz04,ostriker10,benincasa16,ostriker22}. A recent analysis compares these formulae to the pressure calculated directly from a numerical integration of the force equation \citep{corbelli25}. The most accurate expression for a plane-parallel layer was found to be
\begin{equation}
    P={\pi\over 2}G\Sigma_{\rm g}\Sigma_{\rm total}^{\rm g}
\end{equation}
where $\Sigma_{\rm g}$ is the gas surface density and $\Sigma_{\rm total}^{g}$ is the total effective surface density of all matter inside the gas scale height \citep{corbelli25}.  This equation is accurate to within a few percent with or without dark matter and for the wide range of observables in M33. The other equations in the literature exceed the numerically integrated solution by 10\% to 20\%.  A more comprehensive equation would account for the perpendicular components of gravitational attraction to distant objects such as a bulge or dark matter core \citep{patra19,bacchini19,mancera22,wilson19}, or for a surrounding halo pressure, but for the dIrr galaxies considered here, these additional forces should be small compared to the force from gas, stars, and dark matter inside the gas layer. 

Without dark matter in the disk, \cite{elmegreen89} wrote 
\begin{equation}
    \Sigma_{\rm total}^{\rm g}=\Sigma_{\rm g}+{{c_{\rm g}}\over{c_{\rm s}}}\Sigma_{\rm stars}
\label{eq:nodm}
\end{equation}
for total stellar surface density $\Sigma_{\rm s}$ and for gas and star velocity dispersions $c_{\rm g}$ and $c_{\rm s}$. Equation (\ref{eq:nodm}) also gives a value of $P$ that agrees with direct integrations to within a few percent for the wide range of conditions in M33 (i.e., if there were no dark matter). The expression is useful because all quantities are potentially observable when the gas velocity dispersion is adjusted for other forces on the gas from magnetic fields and cosmic ray interactions with those fields.  \cite{wilson19} suggested multiplying the HI velocity dispersion by $\sqrt 1.3$, which means that the magnetic and cosmic ray pressures increase the turbulent pressure by 30\%. This multiplier was also in \cite{corbelli25} and will be used here; thus we take $c_{\rm g}=\sqrt 1.3\sigma_{\rm HI}$ for observed HI dispersion $\sigma_{\rm HI}$. 

With dark matter in the disk, the total surface density becomes
\begin{equation}
    \Sigma_{\rm total}^{\rm g}=\Sigma_{\rm g}+{{c_{\rm g}}\over{c_{\rm s}}}\Sigma_{\rm s}+2\rho_{\rm DM,eff}H_{\rm g}
    \label{eq:sigtot}
\end{equation}
where $\rho_{\rm DM,eff}$ is the effective dark matter density, for which a good approximation is $\rho_{\rm DM,eff}=\rho_{\rm DM}/\tanh(1)=1.32\rho_{\rm DM}$ for real dark matter density, $\rho_{\rm DM}$. The $\tanh(1)$ term is the fraction of the surface density of a one-component, isothermal layer inside one scale height; it compensates for the much larger scale height of dark matter compared to gas and stars \citep[see][]{corbelli25}. The gas disk scale height, $H_{\rm g}$, comes from a similar expression, 
\begin{equation}
    H_{\rm g}={{\Sigma_{\rm g}c_g^2}\over{2P}}
\end{equation}
since the midplane gas density is $P/c_{\rm g}^2$. With this expression for $H_{\rm g}$, equation (\ref{eq:sigtot}) is quadratic in $P$ and has the solution
\begin{equation}
    P={{\pi G \Sigma_{\rm g}^2}\over{4}}\left(1+{{c_{\rm g}\Sigma_{\rm s}}\over{c_{\rm s}\Sigma_{\rm g}}}+ \sqrt{\left[1+{{c_{\rm g}\Sigma_{\rm s}}\over{c_{\rm s}\Sigma_{\rm g}}}\right]^2 + 8{{c_{\rm g}^2\rho_{\rm DM}}\over{\pi G \Sigma_{\rm g}^2}}}\right)
    \label{finalP}.
\end{equation}

\section{Azimuthally Averaged Quantities}
\subsection{Observations}
The star formation rate per unit area correlates with the gas density or surface density, as originally hypothesized by \cite{schmidt59}, and observed by \cite{kennicutt89},  \cite{buat89}, and \cite{kennicutt98}, with modifications for molecular gas by \cite{wong02}, \cite{gao04}, \cite{bigiel08}, \cite{leroy08} and many others \citep[e.g.,][]{kennicutt12,neumann25}.  The usual interpretation is that star formation scales with the amount of gas and some gas-dependent rate, perhaps regulating how fast the gas turns into molecules and collapses into dense cloud cores \citep{mckee07,schinnerer24,franco25}. 

A similar correlation uses the surface densities divided by a scale height to get the volumetric densities $\rho_{\rm SFR}=\Sigma_{\rm SFR}/(2H_{\rm g})$ and $\rho_{\rm g}=\Sigma_{\rm g}/(2H_{\rm g})$.  In a study of dIrrs by \cite{bacchini20}, this volumetric law had a slope of $1.91\pm0.03$, while the Kennicutt-Schmidt (hereafter KS) relation with surface densities had a slope of $3.17^{+0.11}_{-0.10}$. Using similar dIrr data, \cite{elmegreen15} got slopes for the azimuthally-averaged KS relation of $1.76\pm0.08$ for all the annuli combined, and $2.95\pm 2.09$ for the average of the slopes from individual galaxies.  The \SP\ relation is a combination of the KS law with star formation surface density and the volumetric law with $P$ representing gas density. 
 
\begin{figure*}
\begin{center}
\includegraphics[width=12cm]{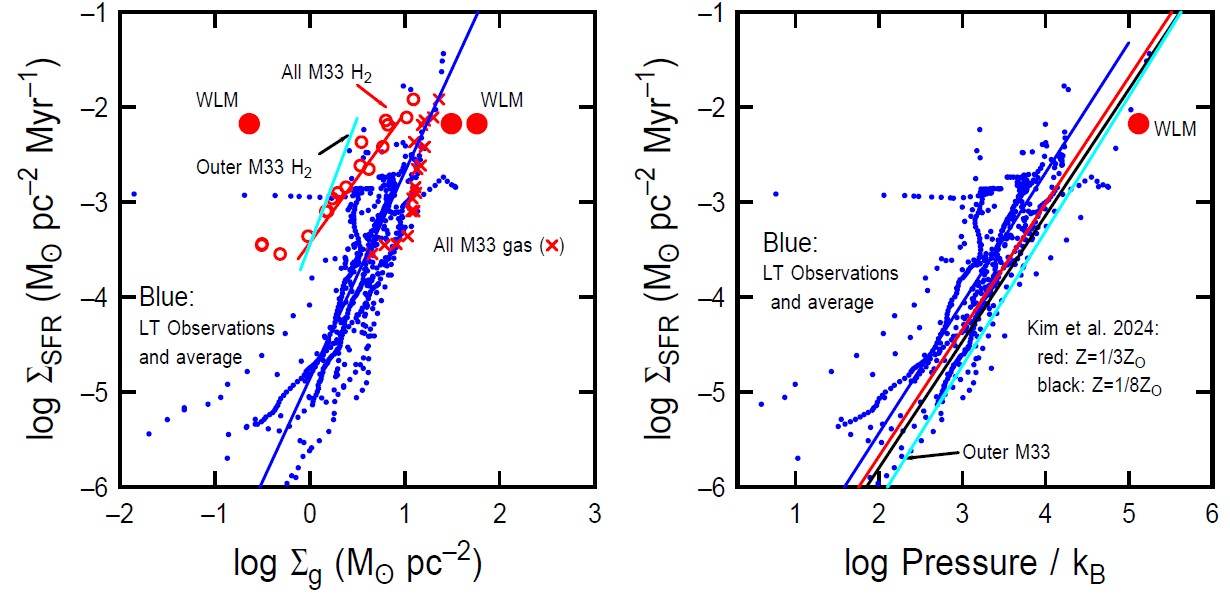}
\caption{The HI KS relation (left) and \SP\ relation (right) for 24 dIrr galaxies (blue dots and blue fitted lines). The fits are given by Eqs. \ref{eq:ks} and \ref{eq:spfit}. Also on the left are the $\Sigma_{\rm SFR}-H_2$ relation for the outer part of M33 (cyan), for all annuli in M33 (red circles and fitting line), and for all gas in M33 (red crosses). The red dots are for WLM with a star formation rate from FUV (like the others), and with gas surface densities for just the CO mass in the star-forming region (left-hand dot), the dark gas in this region (central dot) and the total gas in this region (right-hand dot).  WLM is shown by the red dot in the right-hand panel as well, along with the outer M33 fit in cyan and two models from \cite{kim24}.
}
\label{kssp}
\end{center}
\end{figure*}

Figure \ref{kssp} (left, blue points and blue-line bivariate fit to the correlation) shows the KS relation for azimuthally-averaged quantities measured in annuli of various widths in the LITTLE THINGS galaxies listed in Table 1. The widths of the annuli vary from galaxy to galaxy according to the spatial resolution, from 20 pc in IC 1613 to 340 pc in DDO 52, with a median of 120 pc. The HI pixel size in all cases is $1.5^{\prime\prime}$, chosen to get good sampling of the telescope beams with robust-weighting for best resolution; for a typical 6" beam, that is 4 pixels. The other images were transformed to match the HI pixels. Each string of points in figure \ref{kssp} is from the radial profile of a different galaxy. For all annuli combined, the bivariate fit is
\begin{equation}
\log\Sigma_{\rm SFR}=(2.18\pm0.55)\log\Sigma_{\rm g}- (4.86\pm0.26)
\label{eq:ks}
\end{equation}
for \SSFR\ in $M_\odot$ pc$^{-2}$ Myr$^{-1}$ and $\Sigma_{\rm g}$ from HI in $M_\odot$ pc$^{-2}$ \citep[this is a slightly different sample than in][]{elmegreen15}. The cyan line is the average relationship between \SSFR\ and the molecular surface density in the outer region of M33 \citep[from Table 1 in][]{corbelli25}.  Also added are the annular averages for the $\Sigma_{\rm SFR}-\Sigma_{\rm H2}$ relation in all of M33, as red circles and a red line for the correlation, and the annular averages for the total gas in M33, as red crosses. Three large red points show the star formation rate density for the $18^{\prime\prime}$ region ``A'' in the local dwarf irregular galaxy WLM, with surface densities for the CO-emitting core on the left, the dark-gas in the middle, and the total gas on the right \citep{elmegreen13,rubio15}.

Figure \ref{kssp} (right) shows the correlation between \SSFR\ and the midplane pressure, $P$ (eq. \ref{finalP}) for the dIrrs. The blue points are the observations and the blue line is the bivariate fit,
\begin{equation}
\log\Sigma_{\rm SFR}=(1.37\pm0.29)\log(P/k_{\rm B})-(8.18\pm0.92).
\label{eq:spfit}
\end{equation}
The cyan line is for the outer disk of M33. Two other lines are from the \cite{kim24} simulations: red for a metallicity of 1/3 solar and black for 1/8 solar, chosen as representative for dIrrs. The simulations match the outer disk in M33, where the metallicity is comparable to $1/3$ solar. 

\begin{figure*}
\begin{center}
\includegraphics[width=12cm]{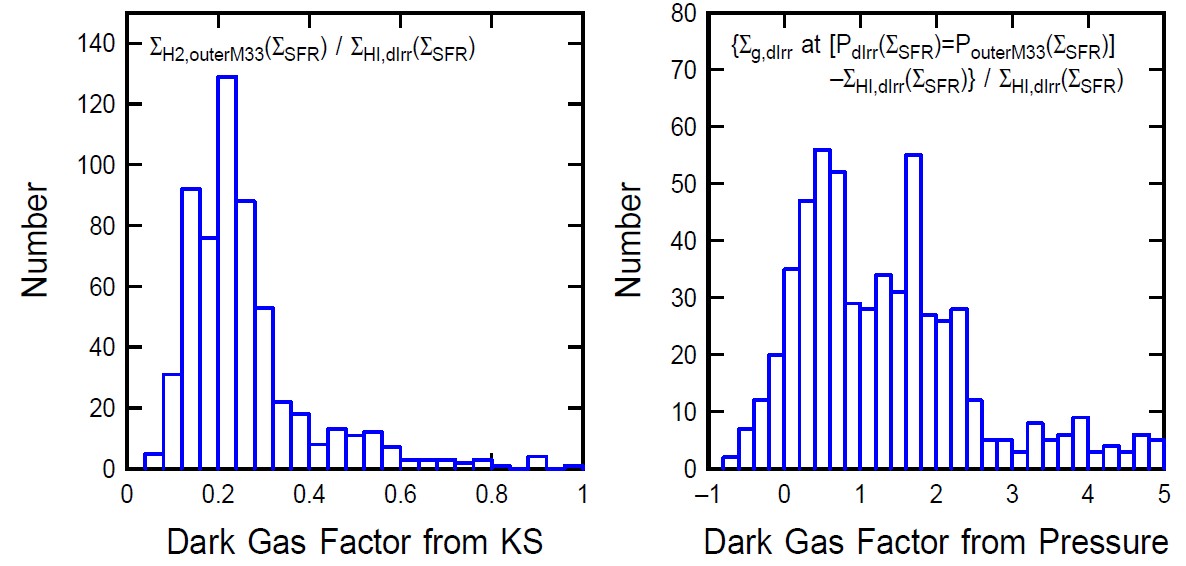}
\caption{Histograms of the dark gas factors, defined to be the ratios of the derived dark gas surface densities to the observed HI surface densities, for each annulus in the dIrrs. On the left, they are calculated from the shift between the outer M33 $\Sigma_{\rm SFR}-H_2$ relationship and the dIrr observations at the same \SSFR. On the right they are calculated by supplementing $\Sigma_{\rm g}$ in  equation (\ref{finalP}) until $P$ for the dIrrs equals $P$ for the outer part of M33 at the same \SSFR. Negative dark gas factors on the right are for dIrr observations at $P$ larger than the M33 pressure. The dark gas fractions from the KS relation are typical for dIrrs, derived by other methods, but some of the dark gas fractions from the \SP\ relation are higher. 
}
\label{histograms}
\end{center}
\end{figure*}

\subsection{Missing gas}
\label{missinggas}
For both panels in Figure \ref{kssp}, the shift between the outer M33 correlation and the dIrr values could indicate missing dark gas in the dIrrs. For the KS relation, the molecular surface density in the outer part of M33 is a factor of $\sim5$ less than the HI surface density in the dIrrs. For all of M33, $H_2$ is about the same factor less than the total surface density at a given \SSFR.  Figure \ref{histograms} (left) shows a histogram of the ratio of the $H_2$ surface density in M33 to the HI surface density in the dIrrs at the same \SSFR. If the dIrrs have the same star formation rate per $H_2$ molecule as the outer part of M33, then $\Sigma_{\rm H2}$ in M33 at a given star formation rate would also be the molecular surface density in the dIrrs, and the plotted ratios in Figure \ref{histograms} would be the molecule-to-atomic gas ratios in the dIrrs. This ratio peaks at $\sim 22$\% with a dispersion of $\pm0.1$. This ratio is about the same as that found in \cite{hunter19} for these dIrrs using the molecular KS relation from \cite{bigiel08}. 

Figure \ref{histograms} (right) shows the relative difference between two quantities. The first is the total gas surface density in a dIrr where there is an observed \SSFR\ that is needed to give the pressure observed in the outer part of M33 at the same \SSFR, which is the cyan line in Fig. \ref{kssp}-right. This total dIrr $\Sigma_{\rm g}$ was found by iteration over $\Sigma_{\rm g}$ until the dIrr pressure equalled the M33 pressure at that \SSFR.  The second quantity is the observed HI gas surface density in the dIrrs, $\Sigma_{\rm HI}$, at that \SSFR. The histogram plots the difference between these two quantities divided by the observed HI  surface density in the dIrr. Because the gas surface density is only one of several contributions to $P$, slight deviations between the dIrr $P$ at a given \SSFR\ and the M33 $P$ at the same \SSFR\ can correspond to a relatively large amount of additional gas needed. The figure suggests some galaxies need more than two times as much gas to bring $P$ up to the value in the outer part of M33. Such large additional gas masses are not likely, given the previous comparisons to molecular gas in the left-hand panel. These extreme gas additions may also be inferred from the scatter in $P$ for a given \SSFR\ on the right-hand side of Figure \ref{kssp}: points far to the left of the cyan line need the most additional gas. For example, a factor of 2 pressure decrease in a dIrr compared to M33 in Figure \ref{kssp}-right corresponds to an additional factor of $\sqrt(2)$ needed for the total gas if there were no other forces contributing to $P$, i.e., no stars or dark matter. 

\subsection{Dark Matter Influence}

Variations in the ratio of dark matter to baryons can affect dIrr pressures much more than analogous variations in spiral galaxies where the dark matter fraction is lower in the main disk.  dIrrs with unusually low dark matter fractions can appear to have low pressures for their \SSFR, distorting possible correlations that may be related to the star formation process.  To check this, we calculated all of the pressures for two dark matter density profiles: the one given by equation \ref{DM1}, which is used for Figure \ref{kssp} and the rest of this paper, and one given by equation (8) in \cite{oh15} with the dark matter parameters in table 2 of that paper.  These two profiles gave similar dark matter densities and pressures for most of the dIrrs, but the Oh et al. profiles have relatively little dark matter compared to equation (\ref{DM1}) for IC 1613 and DDO 50, and NGC 3738 has a much larger $\rho_{\rm DM}$ than the others with the \cite{oh11} profile relative to the gas and stars.  Galaxy inclinations are needed for the Oh et al. values, which are derived from rotation curves, and the inclinations could be overestimated for IC 1613 and DDO 50. Inclinations that are too high make the inclination-corrected rotation speeds smaller than the true rotation speeds \citep{oman16}.  NGC 3738 is different because it is unusually centrally concentrated in optical images, presumably because of a merger with another dwarf \citep{ashley17}.

\begin{figure*}
\begin{center}
\includegraphics[height=8cm]{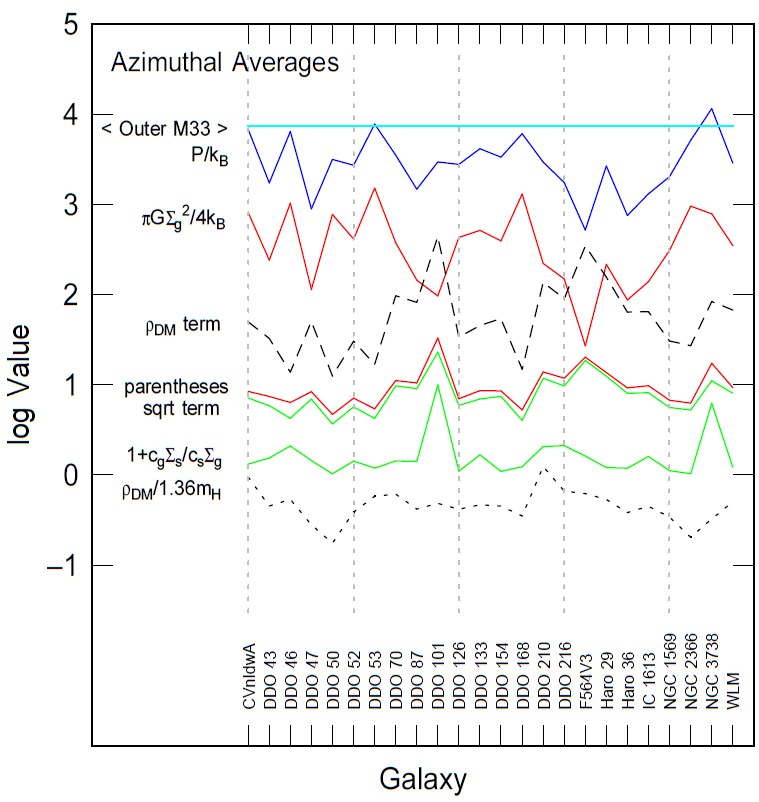}
\includegraphics[height=8cm]{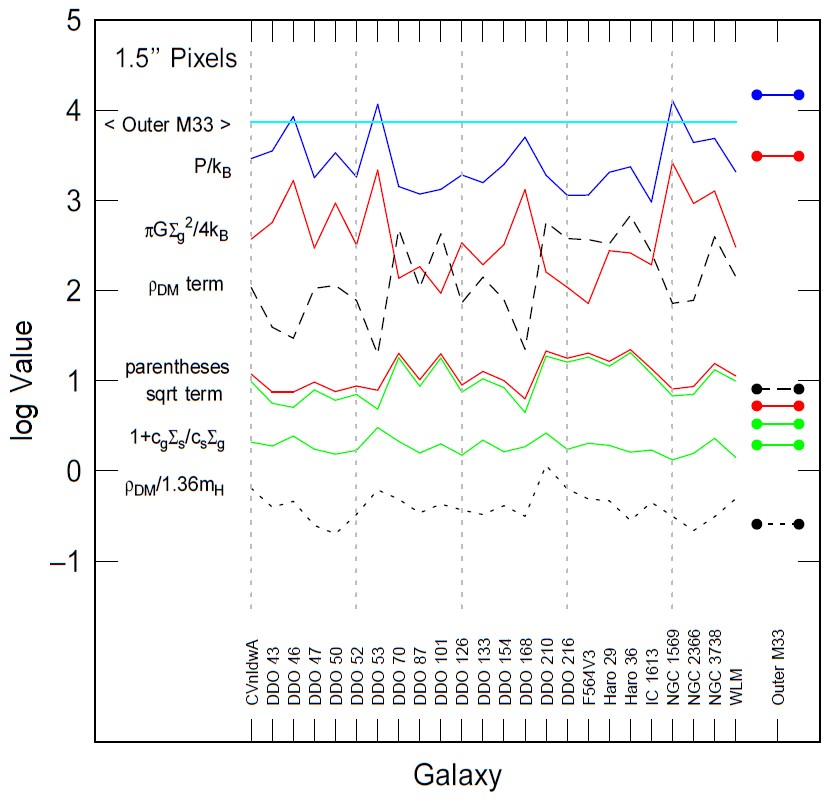}
\caption{The average values for the pressure in the outer part of M33 (cyan line) and the dIrrs (blue curve), and for the terms in the pressure equation for the dIrrs (red, green, black lines), evaluated for \SSFR\ in units of $M_\odot$ pc$^{-2}$ Myr$^{-1}$ between $10^{-3}$ and $10^{-4}$. The pressure variations are relatively small, which is consistent with the good correlation between $\Sigma_{\rm SFR}$ and $P$. High values of $\rho_{\rm DM}$ are compensated by low values of $\Sigma_{\rm g}$. The points on the far right represent values for the outer disk of M33 using the same color scheme. These points for M33 pressure are averages over outer disk pixels which have $\Sigma_{SFR}$ in the same range as considered for dIrrs, while the cyan line for M33 pressure is the value from the fit to the \SP\ relation at $\log \Sigma_{SFR} = -3.5$. }
\label{Pparts}
\end{center}
\end{figure*}

Figure \ref{Pparts} shows the value of $P$ and various terms in the equation for $P$ (eq. \ref{finalP}) for each galaxy, measured as the average of the linear values of these terms in a range of \SSFR\ between $10^{-4}$ and $10^{-3}$ in units of $M_\odot$ pc$^{-2}$ Myr$^{-1}$.  This range is in the middle of the distribution of points in Figure \ref{kssp}.  The left-hand panel is for the azimuthally-averaged quantities and the right-hand panel is for the $1.5^{\prime\prime}$ pixels, discussed more below.  Also in the right-hand panel the average values for the outer disk of M33 are shown, for comparison.

Each panel in Figure \ref{Pparts} plots as a cyan line the pressure for the outer M33 disk at a star formation rate in the middle of this range, $\log\Sigma_{\rm SFR}=-3.5$.  The blue line at the top of the figure is the total $P/k_{\rm B}$, and this is the product of the two red curves, the top one plotting $\pi G \Sigma_{\rm g}^2/4k_{\rm B}$ for the term in front of the parentheses, and the bottom one the term in parentheses. This top red curve plots only the gas quantity so its variation on a log plot like this, at fixed \SSFR, is a measure of the horizontal scatter in the KS relation. Often, the two red curves compensate for each other, in the sense that when one goes up the other goes down, making the blue pressure curve more uniform than the red pure-gas term, and consistent with previous findings that the \SP\ correlation is tighter than the KS correlation \citep{ellison24,corbelli25}. 

Also in Figure \ref{Pparts}, the bottom red curve, which is the term in parentheses, is the sum of the two green curves. The bottom green curve is the sum of the gas and stellar contributions to the total surface density in the gas layer divided by the gas contribution, i.e., $1+c_{\rm g}\Sigma_{\rm s}/c_{\rm s}\Sigma_{\rm g}$.  The top green curve, which is the square root term inside the parentheses, is the additional contribution from dark matter. Note that if $\rho_{\rm DM}=0$, then the top-green curve would be twice the bottom-green curve (shifted up by $\log 2$), but usually the shift is much more than this, and the parentheses term in red is nearly equal to the square root term in green. This near equality shows the high importance of dark matter in $P$. 

The dark matter term itself inside the square root is the dashed black line in Figure \ref{Pparts}, and the dark matter density alone, converted to an equivalent atomic density by dividing by $1.36m_{\rm H}$, is the dotted black line.  The dark matter density does not vary much from galaxy to galaxy because of the constancy of the central dark matter surface density in equation (\ref{DM2}).  This is why the top red and dashed black lines in Figure \ref{Pparts} alternate in opposite ways. The dominance of the $\rho_{\rm DM}$ term inside the square root term, and therefore inside the parentheses, makes this alternating behavior and therefore smooths out the pressure variations. With the $\rho_{\rm DM}$ term, which is $8c_{\rm g}^2\rho_{\rm DM}/(\pi G\Sigma_{\rm g}^2)$, dominating the parentheses in equation (\ref{finalP}), the pressure scales with the first power of $\Sigma_{\rm g}$, rather than the second power as in the prefactor. 

This dominance of the $\rho_{\rm DM}$ term inside the parentheses of Equation (\ref{finalP}) is the result of the high core surface density of dark matter, $63\;M_\odot$ pc$^{-2}$, which is much higher than the gas and star surface densities in dIrrs. Table 2 in \cite{elmegreen15} gives the average central HI and stellar surface densities and radial profiles for 20 dIrrs overlapping the present sample. The table gives the averages as a natural log, so we convert it to base-10 value here: $\log\Sigma_{\rm g,0}=1.0\pm0.3$ and $\log\Sigma_{\rm 0,s}=0.69\pm0.55$ in $M_\odot$ pc$^{-2}$. The central surface density of dark matter is 6.3 times larger than the average central surface density of HI and 90 times higher than the average central surface density of stars.  Also from that table, the radial profiles of $\Sigma_{\rm g}$ and $\Sigma_{\rm s}$ are approximately exponential with scale lengths of $\sim2.2$ and $\sim1.1$ times the V-band surface brightness scale length, so the HI disk drops slower than the stars. 

The outer-M33 values in the right-hand panel of Figure \ref{Pparts} differ from the dIrr values primarily in the approximately $3\times$ larger \Sg, which increases the prefactor in the expression for $P$ (red line) and lowers the $\rho_{\rm DM}$ term in the square root (black dashed line), by a factor of $\sim10$.  It is well known that dIrrs have very low surface densities, so even the outer part of M33 is not a perfect match for comparison.  

Returning now to the topic of Section \ref{missinggas}: the systematic offset between the cyan line for M33 and the blue curve for $P$ in Figure \ref{Pparts} suggest that $P$ might be underestimated in dIrrs because of missing dark gas. This offset amounts to an average factor of 2.66 higher pressure for a given \SSFR\ in the outer part of M33 than in the dIrrs. The galaxy-to-galaxy variations in the dark matter term do not affect this average much. If we consider that the pure-gas term in front of the parentheses in equation (\ref{finalP}) is 10 to 100 times larger than the parentheses term with dark matter (the difference between the two red curves in Fig. \ref{Pparts}), and that primarily this first term will be affected by dark gas, we can estimate the amount of dark gas needed to make up this factor of 2.66. The additional gas surface density needed would be the square root of 2.66, or 1.63, which means that the ratio of dark gas to HI is 0.63. This is slightly larger than that derived above for the KS relation to bring the M33 outer disk $H_2$ in line with the dIrrs. 

\begin{figure*}
\begin{center}
\plottwo{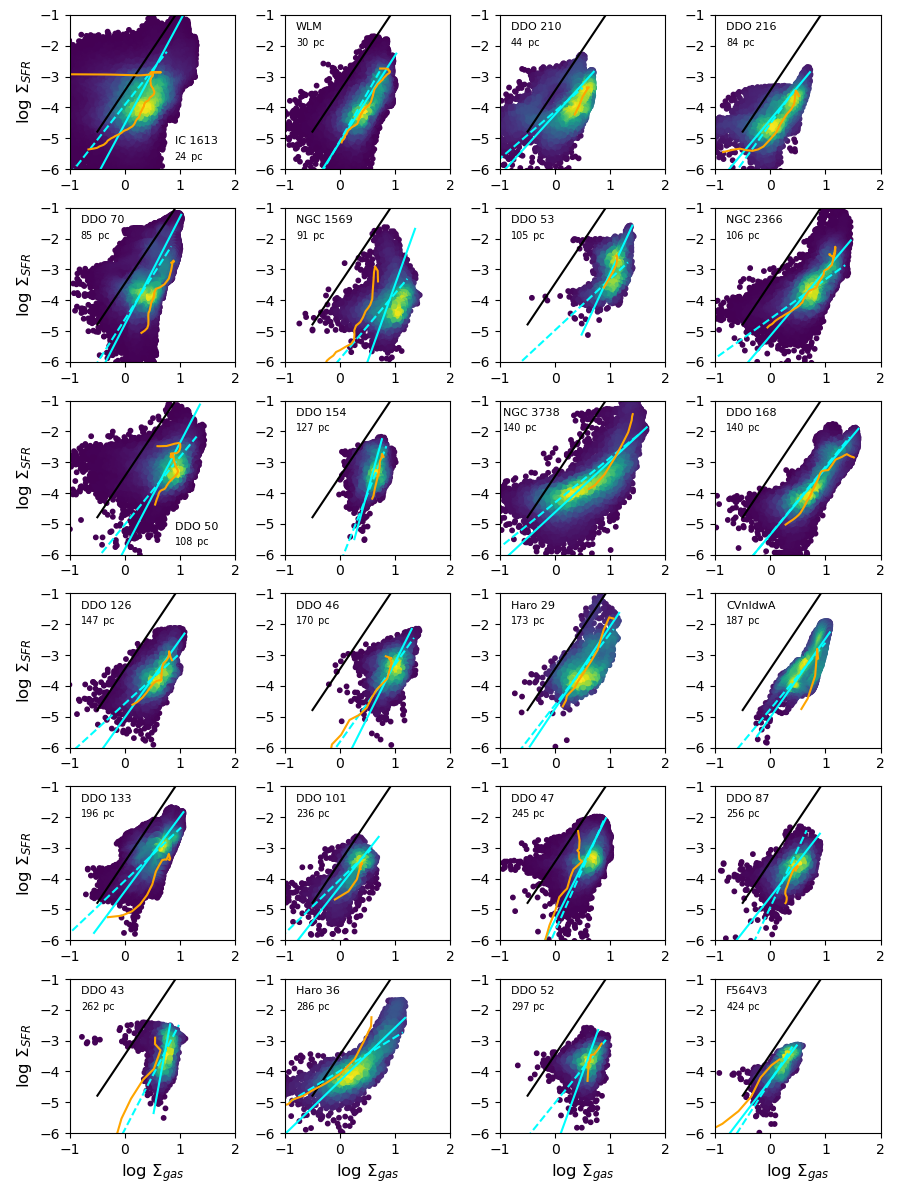}{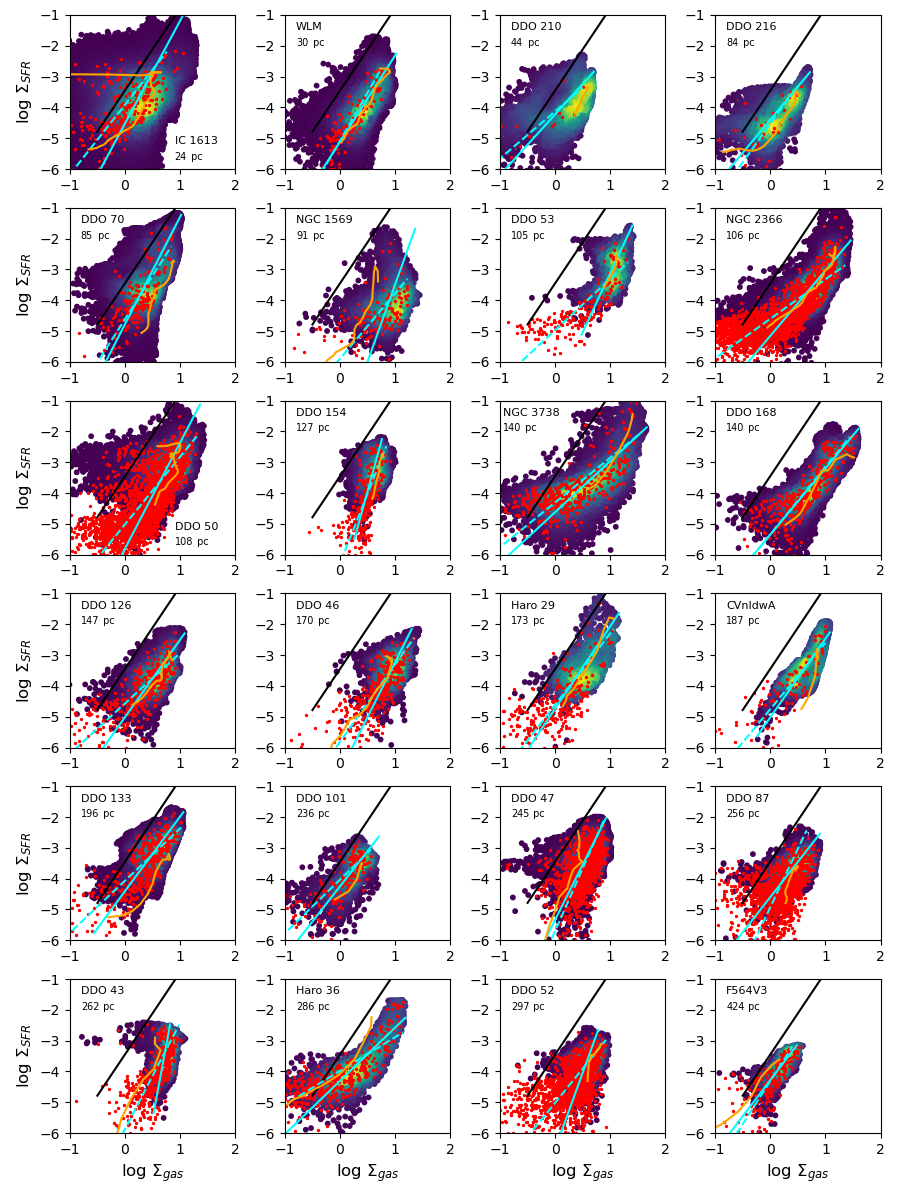}
\caption{Pixel values in the HI KS relation for 24 dIrr galaxies: (left) $1.5^{\prime\prime}$ pixels, (right) the same with 244 pc pixel values superposed in red. The black line is the molecular KS relation for the outer part of M33; the cyan lines are bivariate fits to the data with solid lines for the $1.5^{\prime\prime}$ pixels and dashed lines for the 244 pc pixels; the orange segmented line is the azimuthally-averaged radial distribution from Figure \ref{kssp}. Galaxies are ordered by increasing HI resolution size in pc, as indicated.  
}
\label{SFR-sigma}
\end{center}
\end{figure*}

\section{1.5 arcsec and 244 pc Square Pixels}
\subsection{Observations}

Dwarf irregular galaxies often contain large holes and asymmetries in the HI and star formation \citep{hunter24}, which make azimuthally averaged quantities like those in Figure \ref{kssp} unrepresentative of local processes.  To study the correlations more locally, we evaluate all the relevant quantities for each $1.5^{\prime\prime}$ square pixel in one case, and 244-pc square pixels in another case. Recall that the $1.5^{\prime\prime}$ pixel size originated with the robust-weighted HI maps to get 4 pixels per beam. The pixel size of 244 pc was chosen to match a recent study of pressure and star formation in M33 \citep{corbelli25}. The most distant galaxy, DDO 52, has an HI beam size of $341.2\times259.3$ pc for an average of 297 pc, larger than the 244 pc for this comparison. F564V3 has the largest beam size of 424 pc, and three other galaxies, DDO 43, DDO 87 and Haro 36, come close to 244 pc with average beam sizes of 262 pc, 256 pc and 286 pc.  The other beams are less than 244 pc. We note that the angular beam size of the HI survey is not the same for all galaxies \citep{hunter12}.  The five cases with beams larger than 244 pc should not affect our conclusions because the beam size does not appear to matter for the star formation correlations, which is one of the conclusions of this section.  For dark matter, we continue to use equation (\ref{DM1}), assuming the dark matter is more uniform than the gas, and for $c_{\rm s}$, we continue to use the $M_{\rm B}$-based expression given above, assuming it to be uniform everywhere. 

\begin{figure*}
\begin{center}
\plottwo{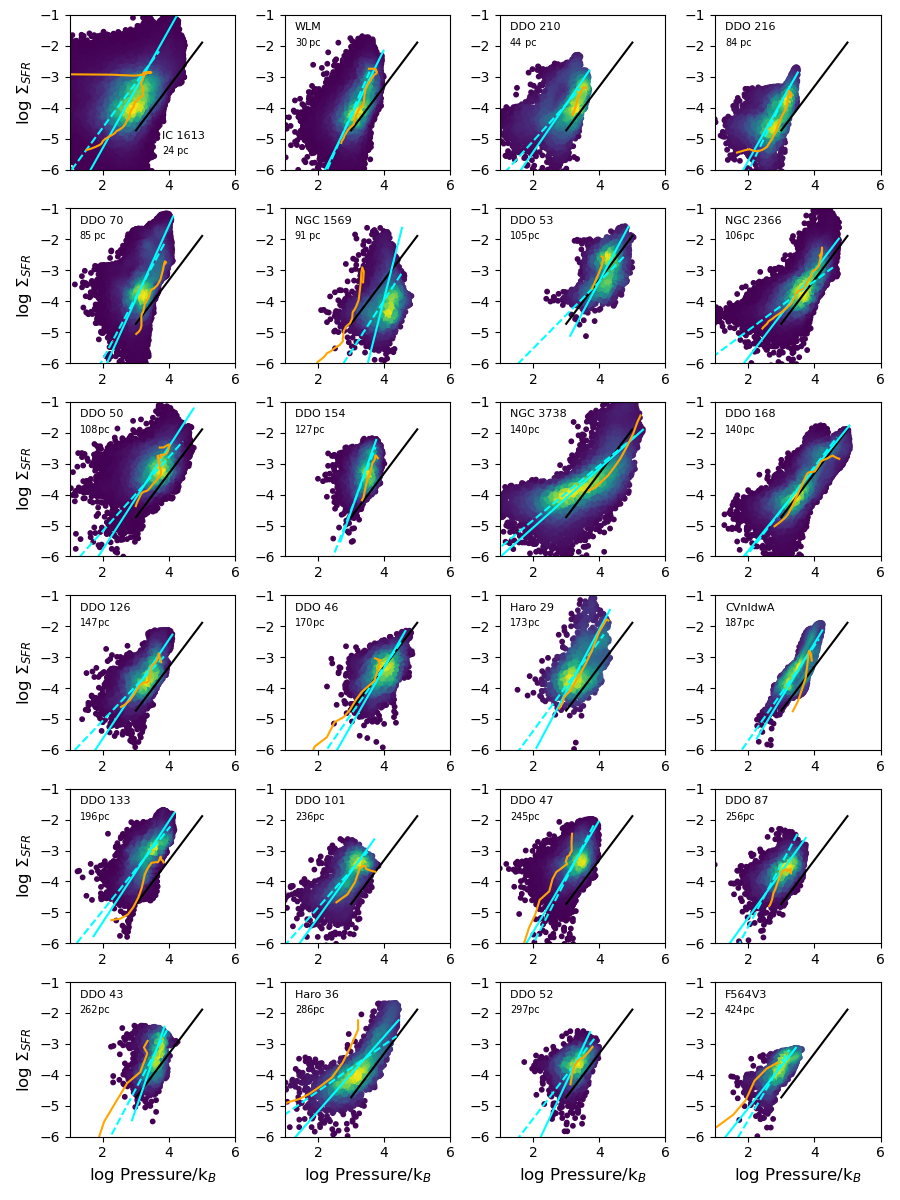}{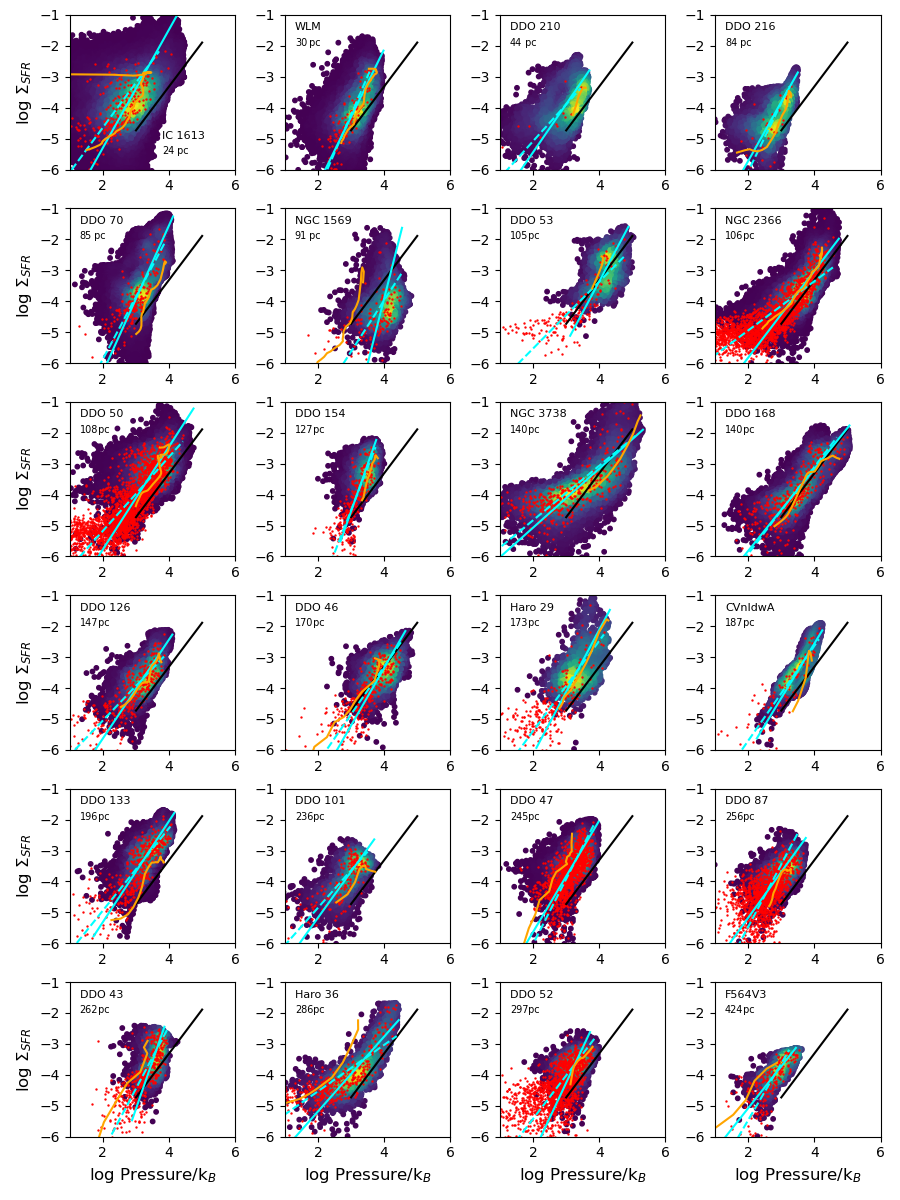}
\caption{Pixel values in the \SP\ relation usinf $1.5^{\prime\prime}$ pixels (left) and 244 pc pixels superposed in red (right). The black line is the molecular KS relation for the outer part of M33; the cyan lines are bivariate fits to the data with solid lines for the $1.5^{\prime\prime}$ pixels and dashed lines for the 244 pc pixels; the orange segmented line is the azimuthally-averaged radial distribution from Figure \ref{kssp}. Galaxies are ordered by increasing HI resolution size in pc, as indicated.  
}
\label{SFR-P}
\end{center}
\end{figure*}

Figure \ref{SFR-sigma} (left) shows the distribution of \SSFR\ versus \Sg\ with $1.5^{\prime\prime}$ pixels. The density of overlapping points is indicated by the color scale.  The order of the galaxies is by increasing physical size of the telescope resolution (proportional to distance), which is indicated in each panel. The resolution is determined by the HI beam and covers between $5^{\prime\prime}$ and $16^{\prime\prime}$. Various lines and curves are overlaid on these points: the black line is the average relationship for the outer part of M33; the solid cyan line is a bivariate least-squares fit to the $1.5^{\prime\prime}$-pixel point distribution, and the orange segmented line is the azimuthally-average radial distribution from Figure \ref{kssp}(left). The main correlations are indicated by the high density regions where the points are yellow. All the galaxies have about the same slope and intercept, as in Figure \ref{kssp}, and they are again shifted to the right of the M33 line.  There is no obvious shift in the distribution with distance (increasing from left to right and top to bottom). Note the range of physical scale, from 24 pc to 424 pc, a factor of 18. Several galaxies have points at low \SSFR\ that trail off to the left. These are HI holes where there is faint FUV emission in the center, presumably from previous star formation. The brightest FUV regions still have a lot of gas. 

Figure \ref{SFR-sigma} (right) shows the same distribution of $1.5^{\prime\prime}$ points but with the 244 pc measurements superposed as red dots.  The 244 pc pixels generally do not have the highest \SSFR\ and \Sg\ values of the $1.5^{\prime\prime}$ pixels because these high values occur when the HI beam is centered on a region of star formation and not diluted by fainter emission within the 244 pc region. The 244 pc points tend to have lower values because of this pixel dilution. The dashed cyan lines in Figure \ref{SFR-sigma} (right) are bivariate fits to the red points. The other lines are as on the left. The two cyan lines are fairly similar except when the 244 pc points trail
off strongly to the lower left. 

Figure \ref{SFR-P} shows the same combination of $1.5^{\prime\prime}$ and 244 pc pixel distributions, but now with \SSFR\ versus $P$. The results are similar to the KS plots in Figure \ref{SFR-sigma} in the sense that the \SP\ relations are nearly independent of spatial resolution from 24 pc to 424 pc.

\begin{figure*}
\begin{center}
\includegraphics[width=12cm]{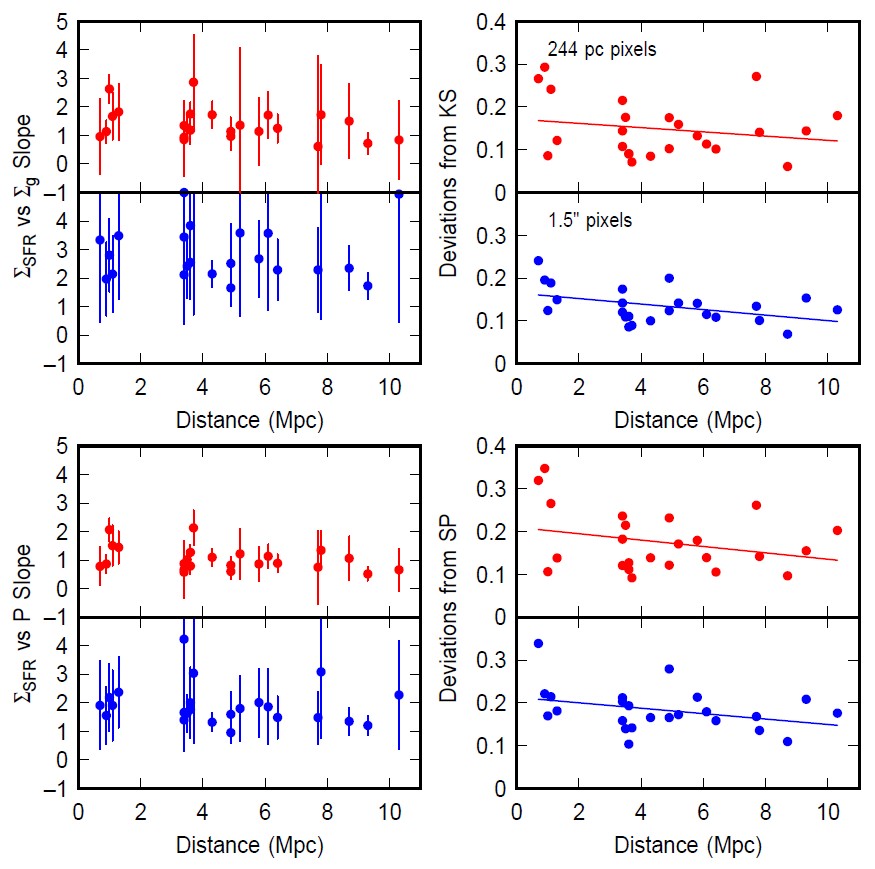}\caption{Slopes over the full range of \SSFR\ (left) and rms deviations around the correlation fits for \SSFR\ between $10^{-4}$ and $10^{-5}\;M_\odot$ pc$^{-2}$ Myr$^{-1}$ (right) for the KS (top) and \SP\ relations (bottom), with blue symbols using $1.5^{\prime\prime}$ pixels and red symbols using 244 pc pixels. 
}
\label{slopes}
\end{center}
\end{figure*}

Figure \ref{slopes}(left) plots all of the slopes for the 24 dIrr galaxies versus distance, using blue points and error bars for the $1.5^{\prime\prime}$ pixels and red points and error bars for the 244 pc pixels. Generally the slopes for both KS and \SP\ are smaller for the 244 pc pixels, as also indicated in by the difference between the dashed and solid cyan lines on the right-hand panels of Figures \ref{SFR-sigma} and \ref{SFR-P}. For KS, the average slopes are $3.3\pm1.8$ and $2.4\pm0.9$ in the $1.5^{\prime\prime}$ and 244 pc cases, respectively, and for \SP\ they are $1.9\pm0.7$ and $1.6\pm0.5$. This slope difference with pixel size is related to the leftward streaks of 244 pc points at low \SSFR\ in Figures \ref{SFR-sigma} and \ref{SFR-P}, which come from HI holes (see above).  

The right-hand side of Figure \ref{slopes} shows the rms deviations perpendicular to the mean fitting lines of the KS and \SP\ relations for the range of \SSFR\ between $10^{-4}$ and $10^{-3}\;M_\odot$ pc$^{-2}$ Myr$^{-1}$ versus distance to the galaxy.  This rms value is the scatter around the trends in Figures \ref{SFR-sigma} and \ref{SFR-P}. The scatter decreases with increasing distance for both the $1.5^{\prime\prime}$ and 244 pc pixels. 
For the KS correlation with $1.5^{\prime\prime}$ pixels, the fit to the scatter points is $(0.17\pm0.03)-(6.46\pm4.92)\times10^{-3}D$ for distance $D$ in Mpc, and with 244 pc pixels, it is $(0.17\pm0.05)-(4.97\pm8.72)\times10^{-3}D$.  For the \SP\ correlation, these fits are $(0.21\pm0.03)-(6.32\pm6.36)\times10^{-3}D$ and $(0.21\pm0.05)-(7.44\pm9.10)\times10^{-3}D$. 
This decrease with $D$ is not surprising for the $1.5^{\prime\prime}$ pixels because the physical size of the measurement gets larger with distance, so more diversity in star formation and cloud structures should be present in the resolution element.  The reason for the negative slope for the 244 pc pixels is not obvious, but the relative error for this slope is larger than for the $1.5^{\prime\prime}$ pixels.

\begin{figure*}
\begin{center}
\includegraphics[width=12cm]{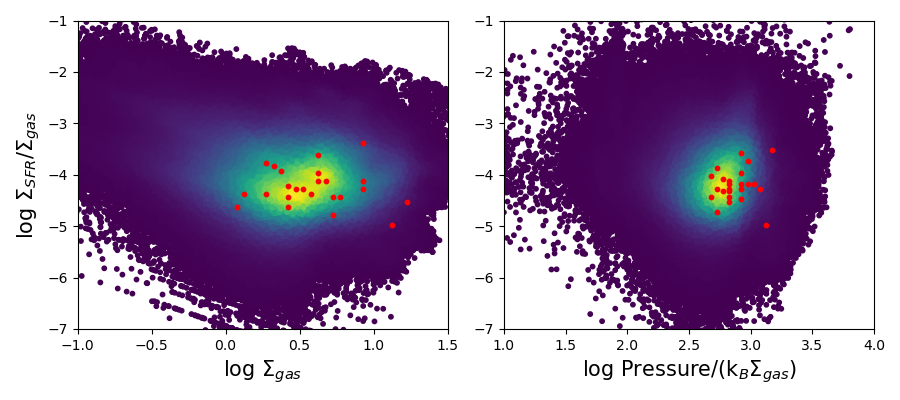}
\caption{
Observed values in the normalized KS (left) and \SP\ (right) distributions with color representing the density of points. The red points are at the positions of peak densities in Figures \ref{SFR-sigma} and \ref{SFR-P}.  The normalization divides $\Sigma_{\rm SFR}$ and $P$ by $\Sigma_{\rm g}$. }
\label{stacked}
\end{center}
\end{figure*}

The similarity of the KS and \SP\ correlations for all of the dIrrs also allows us to meaningfully stack them into single plots. These stacks are shown in Figure \ref{stacked}, but now with both \SSFR\ and $P$ divided by \Sg\ to remove the primary trend with galactocentric radius. Recall that $P/\Sigma_{\rm g}$ is $(\pi/2)G$ times the total mass in the gas layer, and $\Sigma_{\rm SFR}/\Sigma_{\rm g}$ is the inverse of the gas consumption time, which means HI consumption here (even though hidden molecules are actually consumed by star formation). The HI consumption time should be $1/f_{\rm mol}$ times longer than the molecular consumption time for molecule-to-HI ratio $f_{\rm mol}$. In Section \ref{missinggas}, $f_{\rm mol}\sim0.23\pm0.1$, from the shift between the molecular KS relation in the outer part of M33 and the atomic-gas KS relation in dIrrs, where molecules are usually not observed. 

Superposed on Figure \ref{stacked} are red points corresponding to the peak positions in these diagrams for the individual galaxies in Figures \ref{SFR-sigma} and \ref{SFR-P}.  All of the dIrrs have $\log(\Sigma_{\rm SFR}/\Sigma_{\rm g})$ confined to a narrow range of $-4.24\pm0.37$ for the stacked KS relation and $-4.20\pm0.33$ for the stacked \SP\ relation, in units of Myr$^{-1}$. These averages are from the peaks for the individual galaxies. This rate corresponds to an HI consumption time of 16.6 Gyr, nearly independent of $\Sigma_{\rm g}$ or $P/\Sigma_{\rm g}$ near the peaks. Multiplying by our estimated $f_{\rm mol}=0.23$ gives 3.8 Gyr for the molecular consumption time. 

In comparison, the molecular consumption time, $\Sigma_{\rm mol}/\Sigma_{\rm SFR}$, from CO observations of a large sample of galaxies in the PHANGS survey is $10^{3.25\pm0.25}$ Myr in Figure 11 of \cite{schinnerer24}. This is $\sim1.8$ Gyr, smaller by a factor of $\sim2.0$ than the estimated molecular consumption time in dIrrs, which used the outer part of M33 to represent the molecular content in our galaxies given their star formation rates (Section \ref{missinggas}).  The molecular consumption time is also about constant in \cite{schinnerer24} as a function of molecular free-fall time, which is proportional to the inverse square root of average molecular density in the resolved area. Thus, our gas consumption time in dIrrs (when converted to molecules) is within a factor of $\sim2$ of the molecular consumption time for spiral galaxies in the PHANGS survey.

Because both consumption times are independent of which galaxy is considered, and in our case, also independent of physical resolution or distance, the basic star formation law in both spirals and dIrrs is about the same: stars form at an approximately fixed rate per molecule, even at the low gas densities, low stellar mass fractions, and high dark matter fractions in dIrrs compared to spirals.  Of course, we have not answered the main question of what determines the molecular fraction, but only used a molecular fraction in dIrrs scaled from the outer parts of M33, which has physical characteristics similar to those in dIrrs but a more observable CO emission. 

Figure \ref{stacked} (left) shows an increase in $\Sigma_{\rm SFR}/\Sigma_{\rm g}$ with $\Sigma_{\rm g}$ at the positions with the highest point density. That increasing trend has a slope of $\sim2$ which is sensible considering that the slope of the $\Sigma_{\rm SFR}$ vs $\Sigma_{\rm g}$ correlation is $\sim1$ higher, $3.3\pm1.8$ (Fig. \ref{slopes} top-left, blue points).

\begin{figure*}
\begin{center}
\includegraphics[width=12cm]{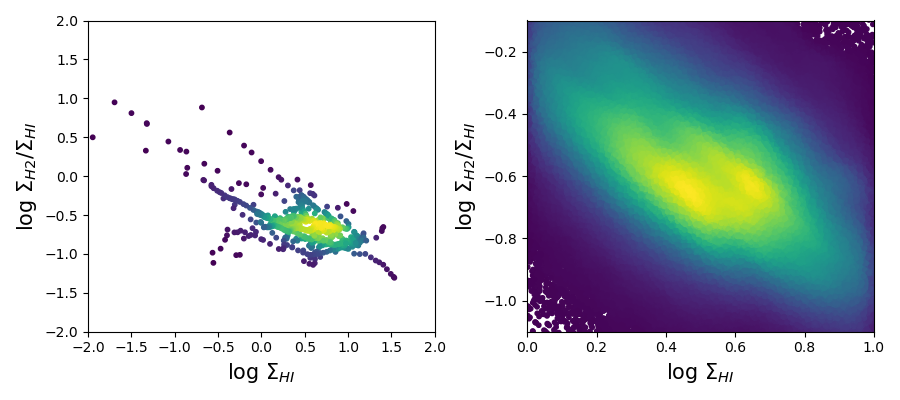}
\caption{
Molecular fraction determined from the ratio of the average molecular surface density in M33 at the \SSFR\ in M33 that equals the \SSFR\ at a particular position in a dIrr galaxy, divided by the HI surface density at that position. The azimuthally-averaged profiles are on the left and the $1.5^{\prime\prime}$ pixels are on the right, showing only a zoom-in of the peak density of points. The molecular fraction is plotted against the dIrr HI surface density. The streaks to the upper left in both plots are from HI holes, where there is residual FUV emission from older star formation, giving the false impression there is H$_2$ present, and little HI gas currently.  Only the regions with peak densities should be considered, and in these regions, the molecular fraction is approximately constant for a range in $\Sigma_{\rm HI}$ that spans a factor of $\sim3$.}
\label{fmol}
\end{center}
\end{figure*}

We consider the molecular fraction again in Figure \ref{fmol} which plots $\Sigma_{\rm H2}/\Sigma_{\rm HI}$ versus $\Sigma_{\rm HI}$. As above, $\Sigma_{\rm H2}$ is determined from the molecular surface density in outer-M33 at the same \SSFR\ as a position or annulus in a dIrr, where $\Sigma_{\rm HI}$ is measured. On the left are the molecular fractions from the azimuthal averages, equivalent to the figure \ref{histograms} histograms but plotted in a different way, and on the right are the $1.5^{\prime\prime}$ pixels.  The regions of highest point density in each plot have a horizontal distribution with a mean value $\log(\Sigma_{\rm H2}/\Sigma_{\rm HI})\sim-0.64\pm0.6$, which corresponds again to $f_{\rm mol}=\Sigma_{\rm H2}/\Sigma_{\rm HI}\sim0.23$ with a factor of $\sim4$ variation among the different galaxies.  The streaks to the upper left in these plots correspond to HI holes where there is FUV from recent star formation but little HI. There were similar streaks at low $\Sigma_{\rm HI}$ in previous plots. 

\section{Discussion}
\label{discussion}

Figures \ref{slopes} and \ref{stacked} indicate that both the KS and \SP\ relations are independent of physical resolution. Recall that all the maps are smoothed to the resolution of the HI, which ranges between $5^{\prime\prime}$ and $16^{\prime\prime}$ and between 24 pc and 424 pc, depending on distance. The $1.5^{\prime\prime}$ pixels are over-resolved, but they allow better centering on peaks and valleys for some of the points than a regular grid at the resolution scale.   The observed independence of our results on physical scale is contrary to the usual assumption that large-scale observations are more representative of average physical processes in a galaxy. While it is true that the scatter in our correlations decreases with increasing scale (Figs. \ref{SFR-sigma}, \ref{SFR-P}), the correlations are the same from scales ranging from 24 pc to 424 pc.  The same scale-independence for the \SP\ correlation was found for M33 between 163 pc and 488 pc \citep{corbelli25}.

This scale independence is reminiscent of the scale independence for SF in general, as shown by power-law two-point correlation functions for cluster position \citep[][and references therein]{zhang01,meena25}, including the youngest clusters which are still embedded in their molecular clouds (Lapeer, private communication). Also, for these youngest clusters, the slope of the two-point correlation function gives a fractal dimension that is about the same as in the general interstellar medium.  For both M33 and the dIrrs in the present study, all of the sampled scales are within this correlation range. 

We also showed that although \SSFR\ is proportional to the HI surface density to a power between $2.4\pm0.9$ and $3.3\pm1.8$ (Fig. \ref{slopes}), when considering all of the galaxies together (Fig. \ref{stacked}), \SSFR\ mostly follows the gas with an average HI consumption time of $10^{1.22\pm0.37}\sim16.6$ Gyr for typical $\Sigma_{\rm g}\sim3\;M_\odot$ pc$^{-2}$ and $P/(k_{\rm B}\Sigma_{\rm g})\sim2.8$ cm$^{3}$K pc$^2M_\odot^{-1}$.  For our $f_{\rm mol}\sim0.22$, this corresponds to 3.7 Myr for the molecular consumption time. 

The range of star formation and gas surface densities in our dIrr data, and the range of pressures, are much lower than in the main parts of spiral galaxies. The pressure is a factor of $\sim3$ to $\sim10$ times lower than even the outer part of M33 (Fig. \ref{Pparts}). Figures \ref{SFR-sigma} and \ref{SFR-P} include \SSFR\ down to $\sim10^{-5}\;M_\odot$ pc$^{-2}$ Myr$^{-1}$, which is as low as we have measured for star formation rates in the deepest images of dIrrs \citep{hunter11,hunter16,hunter25}.  The HI surface densities in Figures \ref{SFR-sigma} and \ref{stacked} concentrate between $\sim1  \;M_\odot$ pc$^{-2}$ and $\sim10\;M_\odot$ pc$^{-2}$. The highest of these values is usually assumed to be where molecules start to appear in spiral galaxies, and it typically lies at the lower end of the KS relation. Star formation in dIrrs, and presumably also molecule formation, is observed to occur at 10 times lower average gas surface densities. 

If the CO-bearing clouds are like CO clouds locally, which require $A_{\rm V}>1$ mag. of extinction \citep[Table 5 in][]{pineda08}, then their individual surface densities will be
\begin{equation}
    \Sigma_{\rm CO-cloud}>20(Z_\odot/Z) \;M_\odot\;{\rm pc}^{-2},
\end{equation}
using $A_{\rm V}=N/(1.87 \times 10^{21}\;{\rm cm}^{-2})>1$ for solar abundance $Z_\odot$ and atomic column density $N$, from \cite{bohlin78}, and assuming a ratio of total-to-selective extinction equal to 3.1.  At the metallicities of our dIrrs ($Z/Z_\odot\sim0.2$ or less), $\Sigma_{\rm CO-cloud}>100\;M_\odot$ pc$^{-2}$, which is $\sim30\times$  higher than the average interstellar surface density. In spiral galaxies, CO cloud surface densities are a factor of $\sim10$ higher than the average interstellar medium. This implies that if star formation requires CO and similar molecules for cooling below cool neutral medium temperatures, then star formation will be confined to relatively smaller, higher-pressure regions in the main disks of dIrrs compared to spirals. The observational consequence is that CO will be highly beam-diluted in surveys of dIrrs, and hard to detect \citep{hunter24}.

The surface density requirement on CO for dIrrs is analogous to the requirement on dense gas tracers like HCN in spiral galaxies.  HCN detection requires $A_{\rm V}\sim8$ mag or more \citep{dame23,schinnerer24}, which is the same cloud surface density as the $A_{\rm V}\sim1$ detection threshold for CO in dIrrs such as WLM where $Z/Z_\odot=1/8$ \citep{rubio15}.  CO in WLM and NGC 6822 is confined to small cores measuring several pc in size which are inside larger clouds observed in HI and dust emission \citep{rubio15,schruba17}.  CO is even more of a dense gas tracer in dIrrs than HCN is in spirals because of the one-tenth lower average pressures in dIrrs. In dIrrs, the CO is surrounded by a lower-density cloud, which emits primarily in [CII] \citep{cigan16,cormier19,madden20,cigan21,ramambason24}, like HCN is surrounded by lower-density clouds in spirals that emit primarily in CO.

Also like HCN in the cores of spiral galaxy CO clouds, CO in the cores of dIrr clouds should be strongly self-gravitating, whereas in spirals, CO-emitting regions can be mostly diffuse \citep{evans22}.  This strong self-gravity in dIrrs is directly inferred from the high internal pressure required for CO, which is the result of the $A_{\rm V}>1$ mag requirement at $Z/Z_\odot\sim0.2$. For this metal abundance where $\Sigma_{\rm CO-cloud}\sim100\;M_\odot$ pc$^{-2}$, the internal cloud pressure is $(\pi/2)G\Sigma_{\rm CO-cloud}^2=3.4\times10^5k_{\rm B}$. This is $\sim100\times$ larger than the average pressure, as shown in Figures \ref{kssp} and \ref{SFR-P}.  Such high pressures require strong self-gravity in the CO regions to hold the clouds together.

\section{Conclusions}
A compilation of \SSFR, \Sg, and $P$ for 24 dIrr galaxies shows correlations similar to those observed in spiral galaxies, even though \Sg\ and $P$ in the dIrrs are lower by factors of $\sim10$ or more. These comparisons were made with azimuthally-averaged radial profiles, and with maps having $1.5^{\prime\prime}$ pixels and 244 pc pixels. This value of 244 pc is what was used for a previous study of M33 \citep{corbelli25}, where the far-outer regions are in some respects analogous to the dIrrs. 

An important property of the dIrrs is that their dark matter dominates the gravity of the inner disk, contributing significantly to $P$. This strong contribution modulates the positional fluctuations in $P$ for various \SSFR, making the \SP\ correlation tighter than without dark matter.  Dark matter makes dIrrs a good test of star formation models based on pressure. 

The $\Sigma_{\rm SFR}-\Sigma_{\rm g}$ and \SP\ correlations are independent of distance and physical resolution from $\sim20$ pc to $\sim400$ pc, suggesting scale-free properties analogous to those for star clusters. The \SP\ correlation was found to be independent of resolution also in M33 for scales ranging from 163 pc to 488 pc \citep{corbelli25}.  The rms dispersions around these correlations decrease with increasing distance and physical resolution for both the dIrrs and M33. This decrease is presumably because each measurement is closer to the average disk property when the resolved spatial scale is larger. 

CO is not observed in most of the dIrrs studied here. To estimate the average ratio of dark or molecular gas to HI surface densities in the dIrrs, we ratioed molecular surface densities in the outer part of M33 to the HI surface densities in dIrrs at the same local \SSFR.  This led to an average ratio of $\sim0.23\pm0.1$, and with this average, a consumption time per molecule of $\sim3.8$ Gyr, which is a factor of $\sim2$ longer than in typical spiral galaxies. 

We considered the nature of CO-emitting clouds in dIrrs, where the metallicity and pressure are low. Assuming the minimum extinction to make a CO region is the same in dIrrs as in the Milky Way, i.e., $A_{\rm V}\sim1$ mag., CO in dIrrs should form only in dense, self-gravitating cloud cores. In this respect, CO in dIrrs is similar to dense gas tracers like HCN in spiral galaxies. The high surface density corresponding to $A_{\rm V}\sim1$ mag at low metallicity, combined with the low interstellar pressures in dIrrs, also implies that CO cores in dIrrs are strongly self-gravitating.

{\it Acknowledgements}
Flagstaff sits at the base of mountains sacred to tribes throughout the region. 
We honor their past, present, and future generations, who have lived here for millennia and will forever call this place home.  EC acknowledges finantial support from IINAF-Mini Grant RF-2023-SHAPES. GALEX data presented in this article were obtained from the Mikulski Archive for Space Telescopes (MAST) at the Space Telescope Science Institute; the DOI for the paper describing the GALEX GR3 data release is 10.1086/520512. The authors are grateful to the referee for a careful reading.

\newpage

\bibliography{refs}{}
\bibliographystyle{aasjournalv7}

\end{document}